\documentclass[conference]{IEEEtran}
\IEEEoverridecommandlockouts

\usepackage{amsmath,amssymb,amsfonts}
\usepackage{accents}
\usepackage{algorithm}
\usepackage{algorithmic}
\usepackage{graphicx}
\usepackage{multicol}
\usepackage{placeins}
\usepackage{textcomp}
\usepackage{xcolor}
\usepackage{subfig}
\usepackage{url}
\usepackage{hyperref}

\usepackage[backend=biber,style=ieee,sorting=none,firstinits=true]{biblatex}
\usepackage[nodisplayskipstretch]{setspace}
\newcommand{\XK}{\boldsymbol{\protect\accentset{\circ}{X}}}
\newcommand{\xK}{\boldsymbol{\protect\accentset{\circ}{x}}}

\IEEEoverridecommandlockouts
\IEEEpubid{\makebox[\columnwidth]{978-1-7281-1895-6/19/\$31.00~\copyright~2023 IEEE
\hfill} \hspace{\columnsep}\makebox[\columnwidth]{ }}

\begin{document}

\title{%
Solving FDR-Controlled Sparse Regression Problems with Five Million Variables on a Laptop
	\thanks{The first author is supported by the federal German BMBF Clusters4Future initiative curATime, within the curAIsig project. The first, and the second author are supported by the LOEWE initiative (Hesse, Germany) within the emergenCITY center. The third author is supported by the ERC Starting Grant ScReeningData under grant number 101042407.
    The final version of this paper is available at \url{https://ieeexplore.ieee.org/document/10403478}
	}
}

\author{\IEEEauthorblockN{Fabian Scheidt, Jasin Machkour, Michael Muma}
\IEEEauthorblockA{\textit{Robust Data Science Group} \\
\textit{Technische Universität Darmstadt}\\
64283 Darmstadt, Germany\\
\{fabian.scheidt, jasin.machkour, michael.muma\}@tu-darmstadt.de}
}

\maketitle
\IEEEpubidadjcol

\begin{abstract}
Currently, there is an urgent demand for scalable multivariate and high-dimensional false discovery rate (FDR)-controlling variable selection methods to ensure the reproducibility of discoveries. However, among existing methods, only the recently proposed Terminating-Random Experiments (\emph{T-Rex}) selector scales to problems with millions of variables, as encountered in, e.g., genomics research. The \emph{T-Rex} selector is a new learning framework based on early terminated random experiments with computer-generated dummy variables. In this work, we propose the \emph{Big T-Rex}, a new implementation of \emph{T-Rex} that drastically reduces its Random Access Memory (RAM) consumption to enable solving FDR-controlled sparse regression problems with millions of variables on a laptop. We incorporate advanced memory-mapping techniques to work with matrices that reside on solid-state drive and two new dummy generation strategies based on permutations of a reference matrix.  Our numerical experiments demonstrate a drastic reduction in memory demand and computation time. We showcase that the \emph{Big T-Rex} can efficiently solve FDR-controlled Lasso-type problems with five million variables on a laptop in thirty minutes. Our work empowers researchers without access to high-performance clusters to make reproducible discoveries in large-scale high-dimensional data. 
\end{abstract}


\section{Introduction}
\label{sec:Introduction}
Ensuring reproducibility in high-dimensional variable selection is essential in many applications (e.g.,
\cite{wasserman2009high,lima2020variable,buniello2019nhgri,tan2014direction,zoubir2018robust,cattivelli2011distributed,kay1993fundamentals}). Applying methods without appropriate statistical guarantees may result in spending valuable time and resources on researching false positives \cite{tandy2018false, kitchin1990avoidance, choi2020identification, macarthur2012face}.
Consider the field of precision medicine, where a small set of reproducible associations among potentially millions \cite{buniello2019nhgri} in large-scale genomics biobanks \cite{sudlow2015uk} forms the foundation for personalized therapy and drug development and the curation of complex polygenetic diseases. 
Follow-up investigations to study functionally related regions on the genome are time and cost-intensive \cite{uffelmann2021genome}.
Therefore, there is a large interest in methods that control the false discovery rate (FDR) \cite{benjamini1995control, benjamini2001control, BarberCandes2015, candes2018panning, barber2019knockoff, machkour2021terminating, machkour2022false, machkour2023false}.
It is the expected percentage of false discoveries among all discoveries: $\mathrm{FDR} =\mathbb{E}[\# \mathrm{False \ discoveries}/ \# \mathrm{Discoveries}]$ and aimed to be controlled at a user-defined level $\alpha \in [0, 1]$.
An additional requirement is a high true positive rate (TPR), i.e., the expected percentage of true discoveries among all true actives: $\mathrm{TPR} = \mathbb{E}[ \# \mathrm{True \ discoveries}/ \# \mathrm{True \ actives}]$.
Unfortunately, classical FDR controlling methods \cite{benjamini1995control, benjamini2001control, BarberCandes2015} do not apply to high-dimensional data, where the number of variables exceeds the number of observations ($p> n$).
Therefore, in recent years, multivariate FDR-controlling methods for high-dimensional data have been proposed \cite{candes2018panning, barber2019knockoff, machkour2021terminating, machkour2022false, machkour2023false}. 
However, among these, only the methods based on the \emph{T-Rex} selector \cite{machkour2021terminating, machkour2023false, TRexSelectorpackage} are scalable to millions of variables (cf. Fig.~1 in \cite{machkour2021terminating}), while the competing methods are limited to thousands of variables, even on high-performance clusters (HPCs). Unfortunately, access to HPCs is costly, and working on an HPC usually requires a tedious initial setup and advanced user knowledge. \emph{Therefore, this work aims to enable ultra-high-dimensional FDR-controlled variable selection on a laptop.}

\emph{Main Contributions:} 
We provide algorithmic solutions to drastically reduce the random access memory (RAM) consumption of the \emph{T-Rex} selector, which is currently the bottleneck limiting its scalability on a laptop. 
In particular, we (i) introduce the \emph{Big T-Rex}, which uses efficient memory mapping strategies that enable working with memory-mapped matrices that reside on Solid State Drive (SSD); (ii) propose two new strategies to generate dummies for the \emph{T-Rex} selector that rely on permutations of a reference dummy matrix instead of creating multiple matrices; (iii) provide numerical experiments that demonstrate a significant reduction in memory demand and computation time, and showcase FDR-controlled variable selection for sparse regression problems with 5,000,000 variables in thirty minutes on a laptop. An open-source implementation of the proposed method will be made  available in the R package TRexSelector on CRAN \cite{TRexSelectorpackage}.

\emph{Organization}: Sec.~\ref{sec:TRex_selector} recapitulates the \emph{T-Rex} selector. Sec.~\ref{sec:bigTRex_selector} presents the proposed \emph{Big T-Rex} implementations and numerically verifies the consistency of the permutation strategies with the \emph{T-Rex} theory. Sec.~\ref{sec:Simulations} presents simulation results on a laptop to demonstrate the efficiency of the proposed implementations.
Sec.~\ref{sec:Conclusions} concludes this work.

\section{The \emph{T-Rex} Selector}
\label{sec:TRex_selector}
\begin{figure}
    		\centering
            \includegraphics[width=\linewidth]{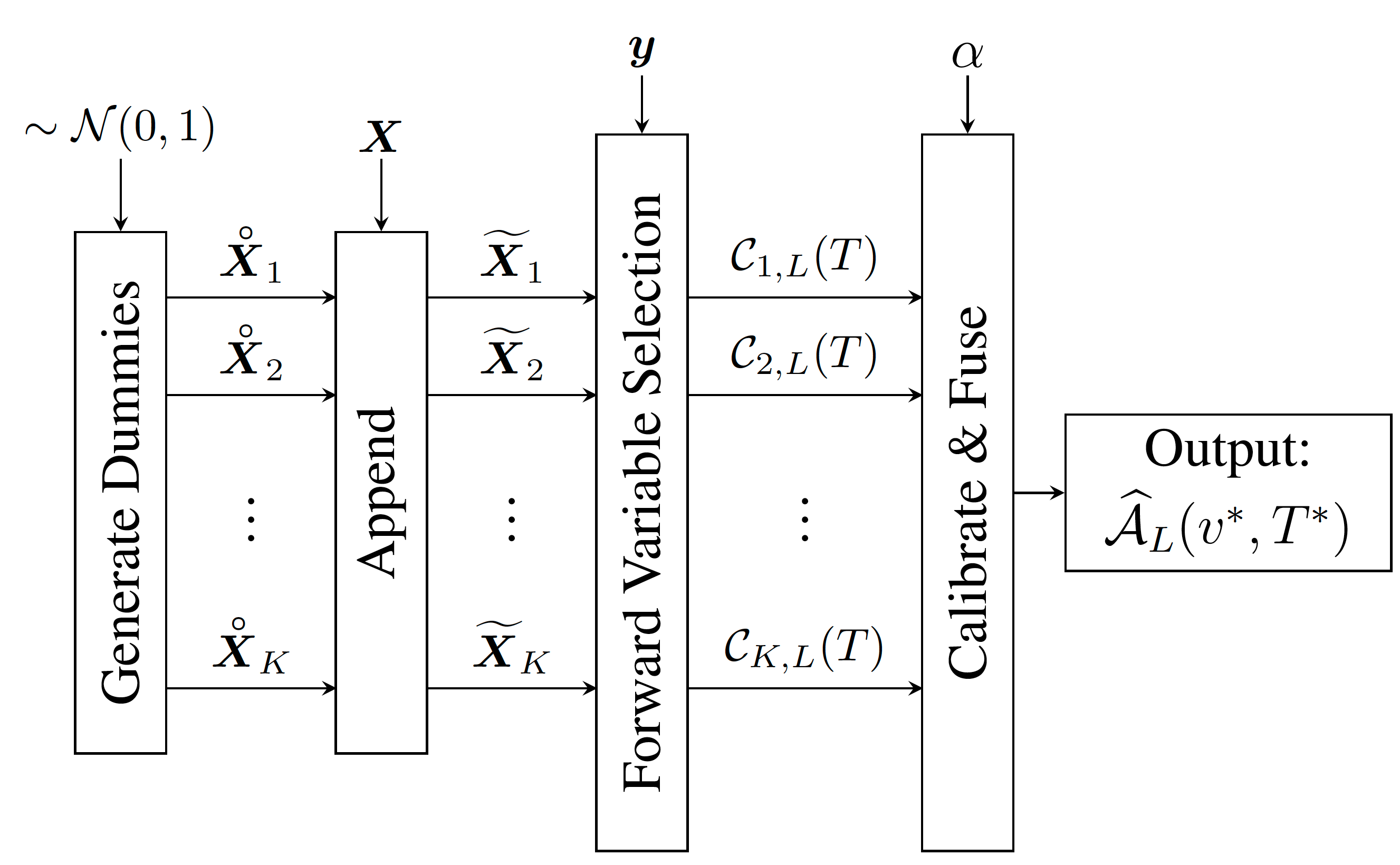}
    	    \caption{Sketch of the \emph{T-Rex} selector \cite{machkour2021terminating}. 
            Ultra-high dimensional problems require storing a set of matrices $\{\widetilde{\boldsymbol{X}}_{k}\}_{k=1}^{K}$ whose data volume usually exceeds a laptop's RAM and thus limits \emph{T-Rex}'s scalability.
            The \emph{Big T-Rex} targets this bottleneck via memory mapping and dummy permutation strategies based on random permutations of a reference matrix.            
            }
         \label{fig:T-Rex_overview}
\end{figure}

The \emph{T-Rex} selector \cite{machkour2021terminating} is a fast FDR-controlling variable selector for low- and high-dimensional data. It controls a user-defined target FDR while maximizing the number of selected variables by mathematically modeling and fusing the solutions of multiple early terminated random experiments where computer-generated dummy variables compete with real variables. 
The \emph{T-Rex} does not require any user-defined tuning. Its inputs are:
\begin{enumerate}
    \item{The predictor matrix $ \boldsymbol{X} = [ \boldsymbol{x}_{1} \, \dots \, \boldsymbol{x}_{p}]$, whose $p$ predictors $\boldsymbol{x}_{1}, \dots, \boldsymbol{x}_{p}$ each contain $n$ samples, i.e., $\boldsymbol{X} \in \mathbb{R}^{n \times p}$. }
    \item{The response vector $\boldsymbol{y} = [y_{1} \, \dots \, y_{n}]^\top$, i.e., $\boldsymbol{y} \in \mathbb{R}^{n}$.}
    \item{The user-defined target FDR level $\alpha \in [0, 1]$.}
\end{enumerate}
A sketch of the \emph{T-Rex} selector algorithm is provided in Fig.~\ref{fig:T-Rex_overview}, and the steps are briefly summarized in the following: \\
\emph{Step~1 (Generate Dummies)}:  A set of $K > 1$ dummy matrices $\{ \XK_{k} = [\xK_{1} \, \dots \, \xK_{L}] \}_{k=1}^{K}$ is generated (each containing $L$ dummies) that can be drawn from a univariate probability distribution with finite mean and variance, e.g., a standard normal distribution (see Theorem~2 of \cite{machkour2021terminating}).\\ 
\emph{Step~2 (Append)}: The dummy matrices are appended to the original predictor matrix $\boldsymbol{X}$ so that a set of enlarged predictor matrices $ \{ \widetilde{\boldsymbol{X}}_{k} = [\boldsymbol{X} \,\, \XK_{k} ] \}_{k=1}^{K}$ is formed, to establish an auxiliary tool for FDR-controlled variable selection in which the original variables compete with the dummies. \\
\emph{Step~3 (Forward Variable Selection)}: A forward variable selection method is applied to each tuple $\{ (\widetilde{\boldsymbol{X}}_{k}, \boldsymbol{y}) \}_{k=1}^{K}$ and terminates for the first time, when $T=1$ dummy variable is included.
For a linear model
\begin{equation}
    \boldsymbol{y} = \boldsymbol{X} \boldsymbol{\beta} + \boldsymbol{\epsilon}
    \label{eq:Eq_Linear_Gaussian_Model}
\end{equation}
with sparse coefficient vector $\boldsymbol{\beta}$ and Gaussian noise $\boldsymbol{\epsilon} \sim \mathcal{N}(\mathbf{0}, \sigma^{2} \mathbf{I})$, the least-angle selector (LARS) \cite{efron2004least} in early terminating step-wise execution mode (T-LARS) \cite{TLARSpackage} is applied.\\
\emph{Step~4 (Calibrate \& Fuse)}:
The outcome is a set of active candidate sets $\{ \mathcal{C}_{k, L}(T) \}_{k=1}^{K}$ from which the $T$ dummies are removed and the relative occurrences of each variable are computed, i.e.,
\begin{equation}
	\Phi_{T, L}(j) = \begin{cases}
		\frac{1}{K} \sum_{k=1}^{K} \mathbb{I}_{k}(j, T, L), &T \geq 1 \\
		0, \ &T = 0
	\end{cases}
	\label{eq:TRex_relative_occurence}
\end{equation}
where the indicator function $\mathbb{I}_{k}(j, T, L)$ equals one if the $j$th variable was selected in the $k$th random experiment.
Then, the false discovery proportion, i.e., a conservative estimate of the proportion of false discoveries among all selected variables, is computed and compared against the user-defined FDR threshold $\alpha$.
If the threshold is not exceeded, the dummy count $T$ is incremented, i.e., $T\leftarrow T+1$, and the forward variable selection is continued until the false discovery proportion exceeds $\alpha$.
Following this procedure, the \emph{T-Rex} selector automatically determines $T$, $L$, and $v$ such that the FDR is controlled at the user-defined target level $\alpha$ (see Theorem~1 in \cite{machkour2021terminating}).\\ 
\emph{Step~5 (Output)}: The active set of variables, i.e., 
\begin{equation}
	\widehat{\mathcal{A}}_{L}(v^{*}, T^{*}) = \{ j : \Phi_{T^{*}, L}(j) > v^{*} \} \ ,
	\label{eq:TRex_estimate}
\end{equation}
where $v^{*}$ and $T^{*}$, respectively denote the optimal values for voting level $v$ and included dummies $T$ that maximize the number of selected variables (see Theorem~3 in \cite{machkour2021terminating}). 

\section{The \emph{Big T-Rex} Implementation}
\label{sec:bigTRex_selector}
\begin{figure}
    \centering
    \includegraphics[width=0.92\linewidth]{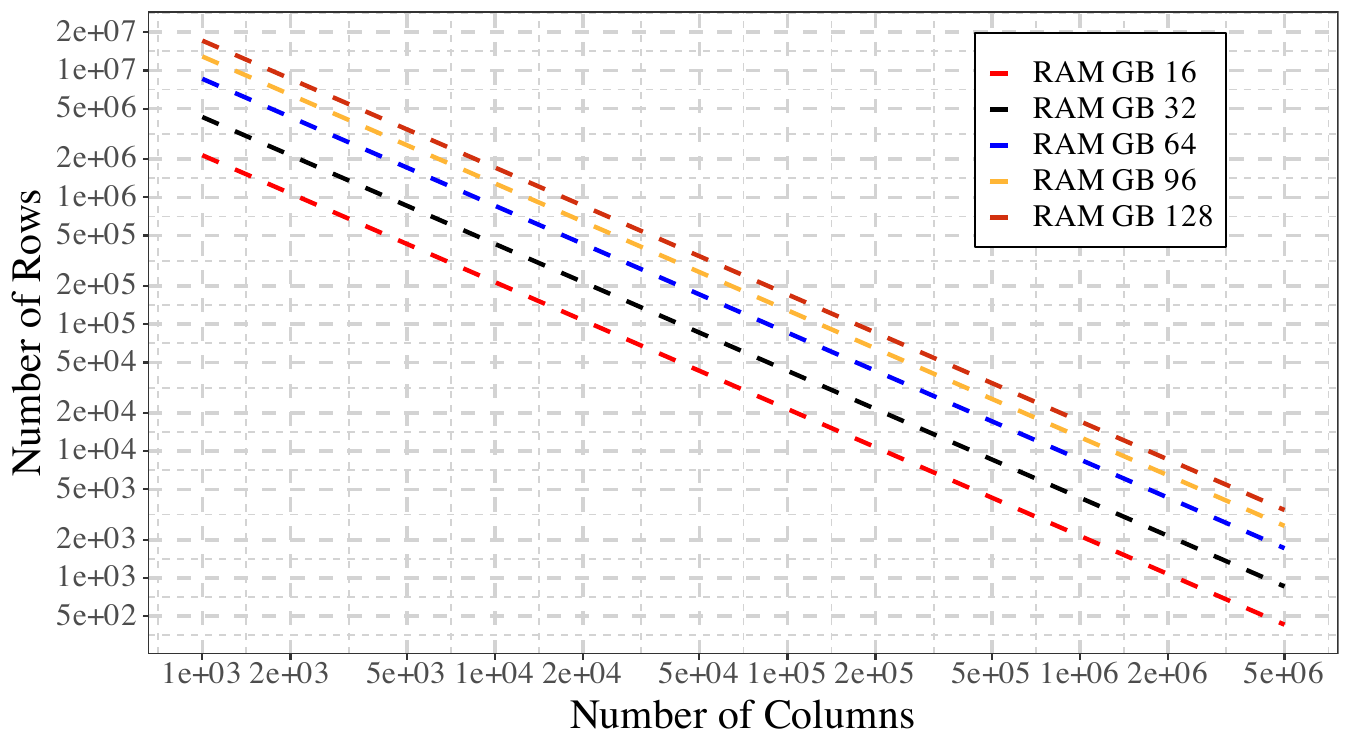}
    \caption{Dimension constraints for given RAM and double-valued matrices.
    Both axes are logarithmically scaled.
    }
    \label{fig:double_Memory_Demand}
\end{figure}

When working with large-scale data on a laptop, RAM becomes a limiting factor. This is numerically illustrated in the following example: A single predictor matrix $\boldsymbol{X}$ of dimensions $p=1,000,000$ and $n=10,000$ consumes 74.5 GB RAM (see Fig.~\ref{fig:double_Memory_Demand}).
The \emph{T-Rex} selector requires storing $K$ enlarged predictor matrices $\{ \widetilde{\boldsymbol{X}}_{k} \}_{k=1}^{K}$.
Thus, even for $L=p$, $K=20$, the resulting set of enlarged predictor matrices consumes 2980 GB RAM, which leads to out-of-memory issues on a laptop.

\subsection{The Big T-Rex Selector}
To compute ultra-high dimensional variable selection problems with the \emph{T-Rex} selector on a laptop, we propose the \emph{Big T-Rex}.
It is a new implementation that drastically reduces the required RAM using memory mapping.
Memory mapping enables inter-process communication and shared file access, such that multiple processes can read and process a single piece of data simultaneously.
These data reside outside of the RAM and can be accessed in an online manner, i.e., loaded and processed sample-wise at demand, without substantial losses in execution time.
The \emph{Big T-Rex} implementation delivers the following major innovations:

 (i) Memory mapping according to \cite{BigMemorypackage} was adopted for the original predictor matrix $\boldsymbol{X}$, and the enlarged predictor matrices $\{ \widetilde{\boldsymbol{X}}_{k} \}_{k=1}^{K}$.
 It allows to store them on SSD memory rather than RAM.
 The data are now processed in an online schedule, one sample at a time.
 For this, row and column indices $i, j \in \mathbb{N}_{0}$ are defined, and within the relevant operations, s.a., data scaling or the $K$ T-LARS executions, $\boldsymbol{X}$ and $\widetilde{\boldsymbol{X}}_{k}$ are sampled element-wise as $\boldsymbol{X}[i,j]$ and $\widetilde{\boldsymbol{X}}_{k}[i,j]$ into the RAM.

(ii) Another form of memory mapping, i.e., serialization as in \cite{Rcerealpackage}, was adopted to store the $k$th T-LARS selector at the end of the $k$th random experiment.
Rather than keeping $K$ T-LARS instances active in RAM, a small set of parameters, comprised of indices and vectors, is mapped on disk for each of the $K$ T-LARS selector instances.
If $T$ is incremented, and another iteration of the $K$ random experiments is computed, the T-LARS instances can be restored by a combination of the stored T-LARS parameters, and the memory-mapped $\widetilde{\boldsymbol{X}}_{k}$, for the $k$th random experiment.

Sec.~\ref{sec:Simulations} verifies, that the \emph{Big T-Rex} produces identical results to \emph{T-Rex} and drastically reduces the RAM consumption.

\subsection{The Dummy Permutating (DP) Big T-Rex Selector}
In the following, a modification of the \emph{Big T-Rex} is proposed, that not only reduces the required RAM, but also the memory consumption on the SSD.
Our key idea is to operate on a single augmented predictor matrix $ \widetilde{\boldsymbol{X}} \in \mathbb{R}^{n \times (p + L)} $, whose dummy part is altered based on permutations of a single dummy reference matrix $ \XK_{\mathrm{ref}} \in \mathbb{R}^{n \times L} $.
Consequently, we avoid storing a set of augmented predictor matrices $ \{ \widetilde{\boldsymbol{X}}_{k} \}_{k=1}^{K} $ on disk.
More specifically, the \emph{DP Big T-Rex} applies a permutation function $\Pi: \mathbb{R}^{n \times L} \rightarrow \mathbb{R}^{n \times L}$ to $ \XK_{\mathrm{ref}} $ and alters the dummy part of $ \widetilde{\boldsymbol{X}} $
by element-wise copies of $ \XK_{\mathrm{ref}} $ as follows:
\begin{align}
	\widetilde{\boldsymbol{X}}_{(1)} &= [\boldsymbol{X} \,\, \XK_{\mathrm{ref}}], &k = 1, \label{eq:perm_dotX_1}\\
	\widetilde{\boldsymbol{X}}_{(k)} &= [\boldsymbol{X} \,\, \Pi_{(k)}(\XK_{\mathrm{ref}})], &k > 1  \label{eq:perm_dotX_k},
\end{align}
where the permutation, for the $k$th random experiment is denoted by the sub-indices in braces.
The realization of the \emph{DP Big T-Rex} is challenging, as a \emph{necessary condition} is to restore the dummy part of $ \widetilde{\boldsymbol{X}}_{(k)} $ in the $k$th random experiment, such that, for a subsequent increment of $T$, the $k$th T-LARS step continues processing with exactly the same data as in the previous iteration.
At first, we select the column index $j = (p+1)$ to start the alteration of $ \widetilde{\boldsymbol{X}}_{(k)} $ from.
For $k > 1$, the algorithm selects the experiment number $k$ as the seed $\theta_{k}$ (i.e., $ \theta_{k} = k$), to control the randomness of a random number generator $\mathcal{G}_{k}$, related to the $k$th random experiment.
Choosing the experiment index as seed generates a deterministic index set $k \in \mathcal{I}_{K} = \{1, 2, \dots, K \} $, which recurs with each increment of $T$, and allows to restore the output of a random generator $ \mathcal{G}_{k} $.
Also for $k>1$, the dummy columns of $ \widetilde{\boldsymbol{X}}_{(k)} $ must be re-standardized after each permutation, to ensure their zero-mean and unit-variance properties.
Subsequently, we present two permutation strategies.
Both operate on deterministic index vectors: 1.) for the rows $ \mathbf{r}_{k} = [1 \ldots n] $ and 2.) for the columns $ \mathbf{c}_{k} = [1 \ldots L] $, for a permutation of $\XK_{\mathrm{ref}}$ within the $k$th random experiment.
Those vectors now become subject to a permutation function $ \pi(\boldsymbol{x} | \theta_{k} ) \, \forall \, k \in \{1, \ldots K\} $, whose source of randomness is controlled by $ \theta_{k} $ and the associated generator $\mathcal{G}_{k}$.

\emph{1.) Permutation Strategy $\mathrm{S}_{1}$:} In strategy $\mathrm{S}_{1}$, the elements of $\mathbf{r}_{k}$ and $\mathbf{c}_{k}$ are randomly shuffled according to
\begin{align}
	\widetilde{\mathbf{r}}_{k} &= \pi(\mathbf{r}_{k}| \theta_{k}) \label{eq:perm_rs}, \\
	\widetilde{\mathbf{c}}_{k} &= \pi(\mathbf{c}_{k}| \theta_{k}) \label{eq:perm_cs}.
\end{align}
Thus, the dummy part of $ \widetilde{\boldsymbol{X}}_{(k)} $, in indices $\mathbf{r}_{k}$ and $p + \mathbf{c}_{k}$, can be reproducibly altered by online loaded element-wise copies of $\XK_{ \mathrm{ref} } $, according to sampling indices $\widetilde{\mathbf{r}}_{k}$ and $\widetilde{\mathbf{c}}_{k}$, as
\begin{align}
	\Pi_{(k)}( \XK_{\mathrm{ref}} ) &= \XK_{\mathrm{ref}}[\widetilde{\mathbf{r}}_{k} , \widetilde{\mathbf{c}}_{k}], \label{eq:Perm_mechanism}\\
	\widetilde{ \mathbf{X} }_{(k)} [\mathbf{r}_{k}, p + \mathbf{c}_{k}] &= \Pi_{(k)}( \XK_{\mathrm{ref}} ). \label{eq:Perm_finish}
\end{align}
In $\mathrm{S}_{1}$ the column index is followed by the row index in looping (column-wise lead).

\emph{2.) Permutation Strategy $\mathrm{S}_{2}$:} Our second strategy increases the randomness by permuting the columns of $\XK_{\mathrm{ref}}$ row-wise.
At first, $\mathbf{r}_{k}$ is randomly shuffled according to (\ref{eq:perm_rs}) based on $\theta_{k}$ as a source of a generator $\mathcal{G}_{k}$.
Using seed $\theta_{k}$, we draw a vector of random seeds $\boldsymbol{\gamma}_{k} \sim \mathrm{ceiling}(\mathcal{U}(0, n))$, where the $i$th element represents the random seed for row $i \in [1, n]$'s column permutation.
These column permutations are reproducibly realized by seeding the generator $\mathcal{G}_{k}$ with the $i$th entry of $\boldsymbol{\gamma}_{k}$ (i.e., $\mathcal{G}_{k}(\boldsymbol{\gamma}_{k}[i])$) to permute $\mathbf{c}_{k}$ according to (\ref{eq:perm_cs}).
The final result is obtained from (\ref{eq:Perm_mechanism}) and (\ref{eq:Perm_finish}), where in looping the rows are sampled first and the column indices follow (row-wise lead).
Algorithm \ref{alg:Processing_Strategies} summarizes both strategies.
\begin{algorithm}
	\caption{Dummy Permutation \emph{Big T-Rex} Strategies}
	\begin{algorithmic}[t]
		\STATE{1. \textbf{Input}: $\widetilde{\boldsymbol{X}}_{(k)} = \left[\boldsymbol{X} \,\, \XK_{(k)} \right]$, $\XK_{\mathrm{ref}}$, $\mathrm{S}_{\mathrm{choice}} \in \{ \mathrm{S}_{1}, \mathrm{S}_{2} \}$.
        }
        \STATE{
        2. \textbf{For} $k = \{2, \dots, K\}$ \textbf{do}:
            \begin{itemize}
                \item[]{
                    \begin{enumerate}
                        \item[i.]{create row index vector $ \mathbf{r}_{k} = [1 \ldots n] $.}
                        \item[ii.]{create column index vector $\mathbf{c}_{k} = [1 \ldots L]$.}
                        \item[iii.]{set seed $\theta_{k} = k$ for random number generator $\mathcal{G}_{k}$.}
                    \end{enumerate}
                }
            \end{itemize}
            \begin{enumerate}
                \item[2.1]{\textbf{IF} $\mathrm{S}_{\mathrm{choice}} == \mathrm{S}_{1}$ invoke permutation strategy $\mathrm{S}_{1}$:
                        \begin{itemize}
                            \item[i.]{random shuffle $\mathbf{r}_{k}$ and $\mathbf{c}_{k}$ using Eqs.~\eqref{eq:perm_rs} and \eqref{eq:perm_cs}.
                            }
                            \item[ii.]{ $\Pi_{(k)}(\XK_{\mathrm{ref}}) = \XK_{\mathrm{ref}}[\widetilde{\mathbf{r}}_{k}, \widetilde{\mathbf{c}}_{k}] $  (column-wise lead).}
                        \end{itemize}}
                \item[2.2]{\textbf{ELSE} invoke permutation strategy $\mathrm{S}_{2}$:
                    \begin{itemize}
                        \item[i.]{generate a vector of seeds for row-wise column permutation $\boldsymbol{\gamma}_{k} \sim \mathrm{ceiling}(\mathcal{U}(0, n))$.}
                        \item[ii.]{random shuffle $\widetilde{\mathbf{r}}_{k}$ using Eq.\eqref{eq:perm_rs}.}
                        \item[iii.]{\textbf{For} $i =\{1, \ldots, n\}$ \textbf{do}:
                            \begin{itemize}
                                \item[a.]{initialize $\mathcal{G}_{k}(\boldsymbol{\gamma}_{k}[i])$.}
                                \item[b.]{random shuffle $\widetilde{\mathbf{c}}_{k}$ using Eq.~\eqref{eq:perm_cs}.}
                            \end{itemize}
                        }
                        \item[iv.]{$\Pi_{(k)}(\XK_{\mathrm{ref}}) = \XK_{\mathrm{ref}}[\widetilde{\mathbf{r}}_{k}, \widetilde{\mathbf{c}}_{k}] $ (row-wise lead).}
                    \end{itemize}
                }
                \item[2.3]{Standardize dummy columns of $\widetilde{\boldsymbol{X}}_{(k)}$.}
            \end{enumerate}
        }
        \STATE{
        3. \textbf{Output}:
            $\widetilde{\boldsymbol{X}}_{(k)}$, $k=1,\ldots,K$, continue \emph{T-Rex} 
            (Fig.~\ref{fig:T-Rex_overview}).
        }
	\end{algorithmic}
	\label{alg:Processing_Strategies}
\end{algorithm}
\subsection{Numerical Analysis of the Dummy Permutation Strategies}
The dummy generation theorem (Theorem~2 in \cite{machkour2021terminating}) demands, that dummies are $\mathrm{i.i.d.}$ realizations following any univariate probability distribution with finite mean and variance.
In the original \emph{T-Rex} implementation, $K$ $\mathrm{i.i.d.}$ dummy matrices were created from a $\mathcal{N}(0, 1)$ distribution. This section provides exemplary numerical evidence that the memory-efficient dummy generation by random permutation of the elements of the reference matrix $ \XK_{\mathrm{ref}} $ via  Eq.~\eqref{eq:Perm_mechanism} with permutation strategies $\mathrm{S}_{1}$ and $\mathrm{S}_{2}$ followed by re-scaling of the columns provides dummies that still follow these conditions.
\begin{figure}[b]
    		\centering
       \subfloat[Permutation strategy $\mathrm{S}_{1}.$\label{fig:QQP_SST1}]{%
            \includegraphics[width=.5\linewidth, height=4.2cm]{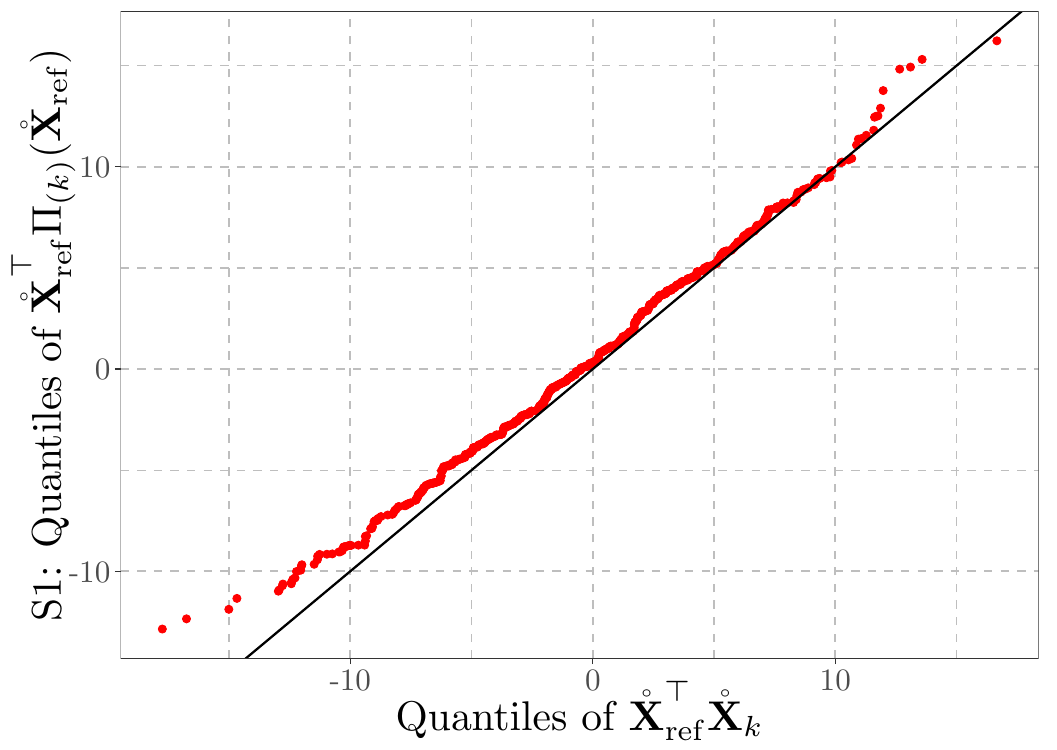}
            }    
        \subfloat[Permutation strategy $\mathrm{S}_{2}.$\label{fig:QQP_SST2}]{%
            \includegraphics[width=.5\linewidth, height=4.2cm]{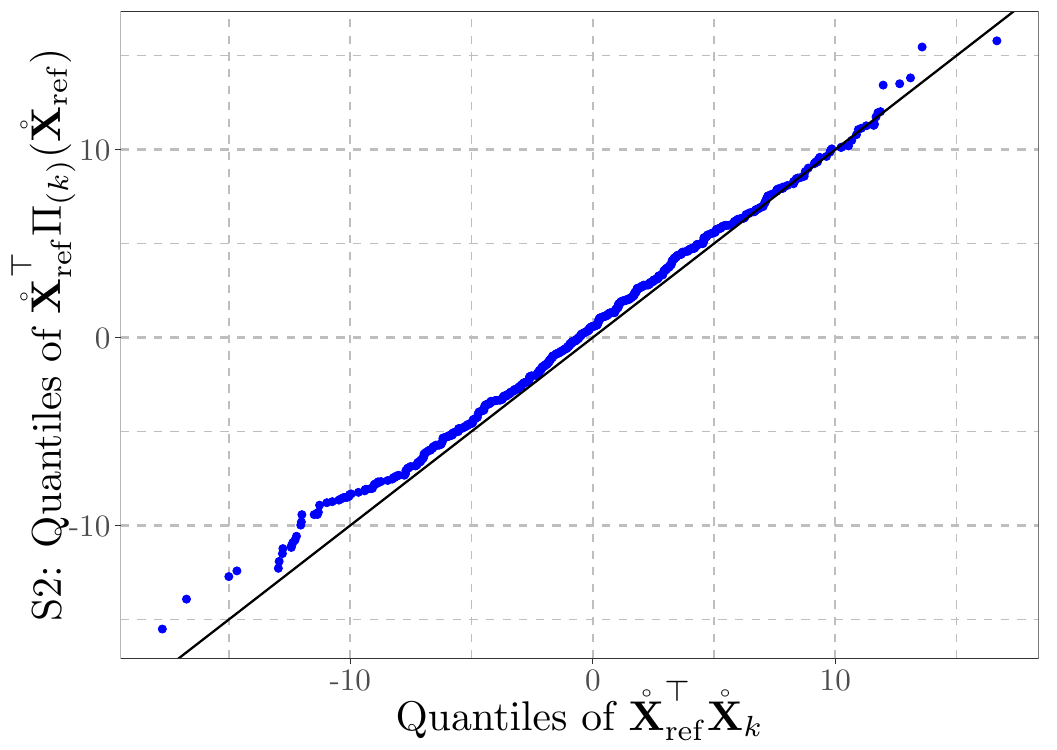}
            }   
    	    \caption{QQ-Plot of the quantiles of $\XK_{\mathrm{ref}}^\top \Pi_{(k)}(\XK_{\mathrm{ref}})$ as used in the \emph{DP Big T-Rex} vs$\mathrm{.}$ those of $\XK_{\mathrm{ref}}^\top \XK_{k}$ as used in the \emph{T-Rex}.}
\end{figure}
In particular, to illustrate that the difference in the dummy generation does not affect the $\mathrm{i.i.d.}$ condition across random experiments, Figs.~\ref{fig:QQP_SST1} and \ref{fig:QQP_SST2} show the QQ-Plots which compare the quantiles of $\XK_{\mathrm{ref}}^\top \Pi_{(k)}(\XK_{\mathrm{ref}})$ to those of $\XK_{\mathrm{ref}}^\top \XK_{k}$, where  $\XK_{\mathrm{ref}}$ and $\XK_k$ are both from an $\mathrm{i.i.d.}$ $\mathcal{N}(0, 1)$ distribution, as in the original \emph{T-Rex}.
Figs.~\ref{fig:QQP_SST1} and \ref{fig:QQP_SST2} confirm that the quantiles are very well aligned and we can conclude that the sampling permutation schemes generate valid dummies for the \emph{T-Rex} selector that allow for FDR-controlled variable selection (see Sec.~\ref{sec:Simulations}).

\section{Simulation Results}
\label{sec:Simulations}
This section presents simulation results comparing the implementation performances of the original \emph{T-Rex} with the proposed: \emph{Big T-Rex} and \emph{DP Big T-Rex} (Algorithm 1).
%
%
The data model for all our simulations is the linear Gaussian regression model (\ref{eq:Eq_Linear_Gaussian_Model}), with $p_1= 10$ true active variables (for an extensive evaluation of the \emph{T-Rex} on other models, see \cite{machkour2021terminating}). 
%
%
The chosen computing platform was an Apple MacBook Pro with M2 Max processor, 64 GB RAM, and a 4 TB SSD with virtual memory support if RAM is exceeded.
Similar setups are commonly used, e.g., by genomics researchers.
All results were obtained via serial processing, i.e., no parallelism was employed.
\subsubsection{Signal-to-Noise Ratio}
The first setup evaluates the FDR and TPR performance in the presence of varying signal-to-noise ratios (SNR).
All implementations were evaluated and averaged based on $400$ Monte Carlo trials.
The data sets were created with $ n = 300 $ observations and $ p = 2,000 $ variables and $\mathrm{SNR} = \mathrm{Var} \left[\boldsymbol{X} \boldsymbol{\beta} \right]/\sigma^{2}$ was realized with $\mathrm{SNR} = \{0.1, 0.25, 0.5, 1, 2, 5, 10\}$.
Also, the target FDR (tFDR) was set to $\alpha = 10\%$.
As discussed in \cite{machkour2021terminating}, the default values of $ K = 20 $ random experiments and $ L = 10p$ were used. 
\begin{figure}[hbp]
    \centering
    \subfloat[FDR-control property (Target FDR $\alpha= 10\%$).\label{fig:fdp_tfdr_01_mcs_data_vs_snr}]{
        \includegraphics[width=0.85\linewidth]{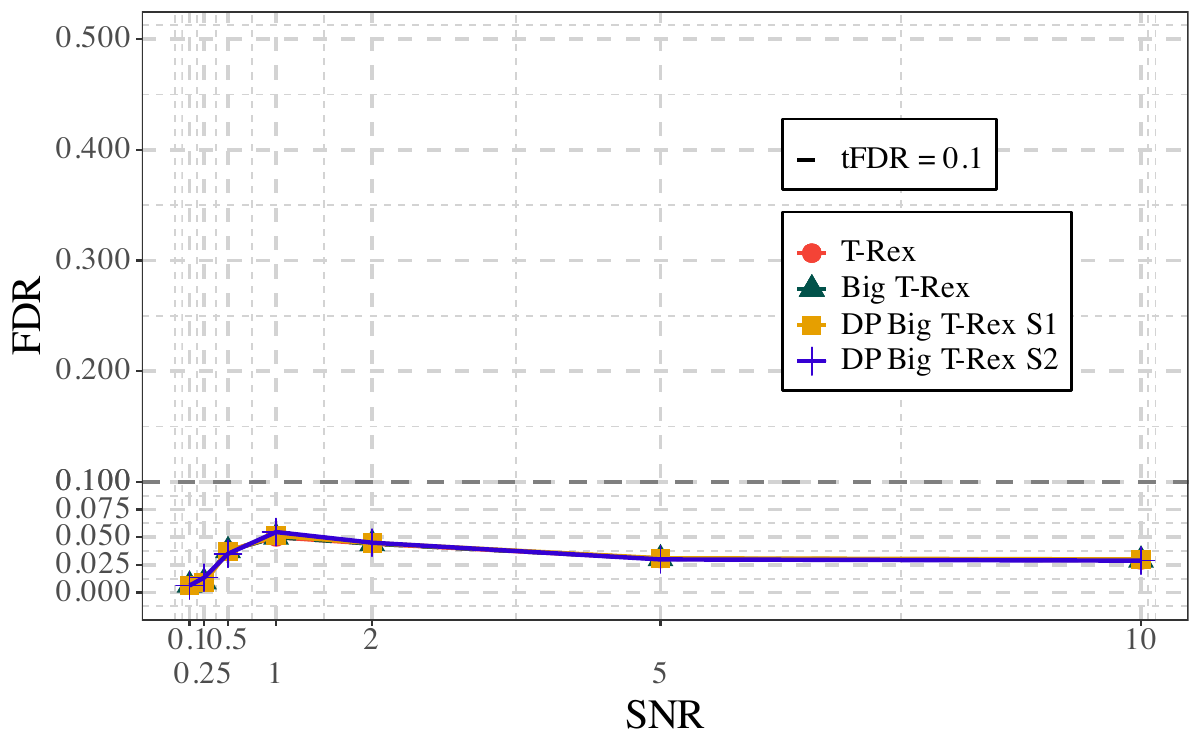}
    }
    \\
    \centering
    \subfloat[TPR performance.\label{fig:tpp_tfdr_01_mcs_data_vs_snr}]{
  	     \includegraphics[width=0.85\linewidth]{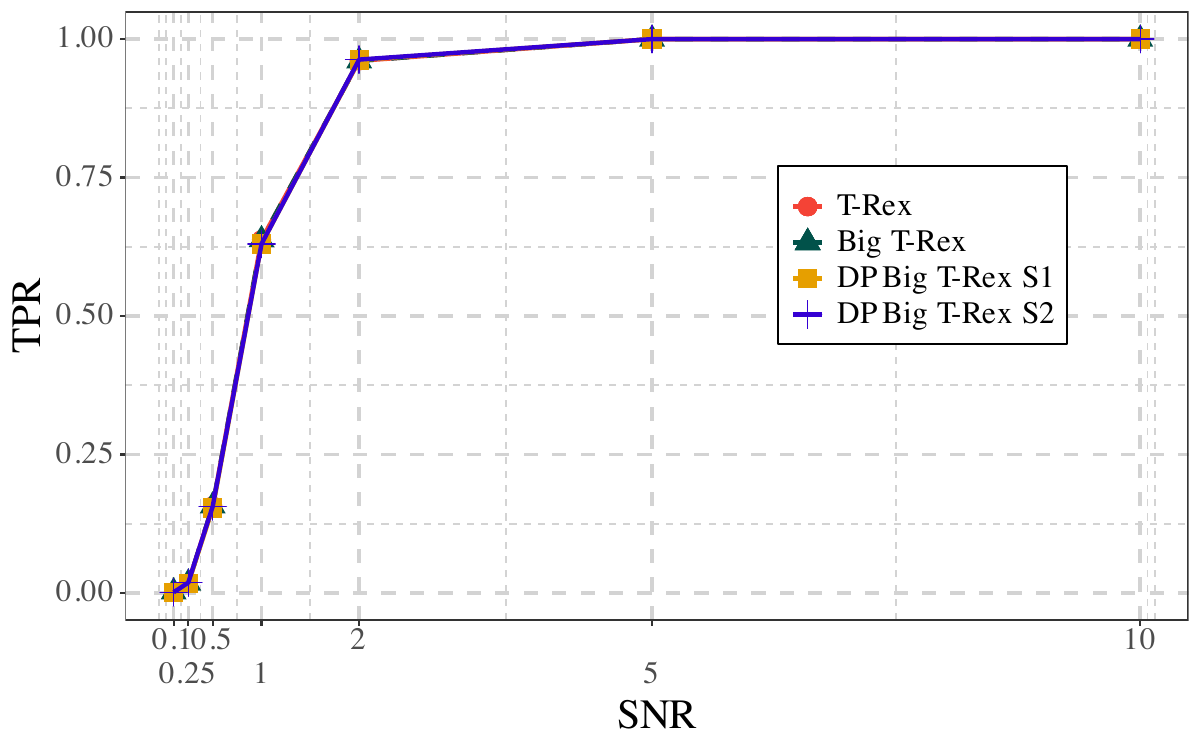}
    }
    \caption{FDR and TPR results for varying SNR.}
\end{figure}
Figs.~\ref{fig:fdp_tfdr_01_mcs_data_vs_snr} and \ref{fig:tpp_tfdr_01_mcs_data_vs_snr} confirm that all variants control the FDR, while as expected \emph{T-Rex} and \emph{Big T-Rex} deliver identical results and the \emph{DP Big T-Rex} delivers almost equal results.

\subsubsection{Memory Demand and Computation Time}
\begin{figure}[htp]
    \centering
    \includegraphics[width=0.85\linewidth]{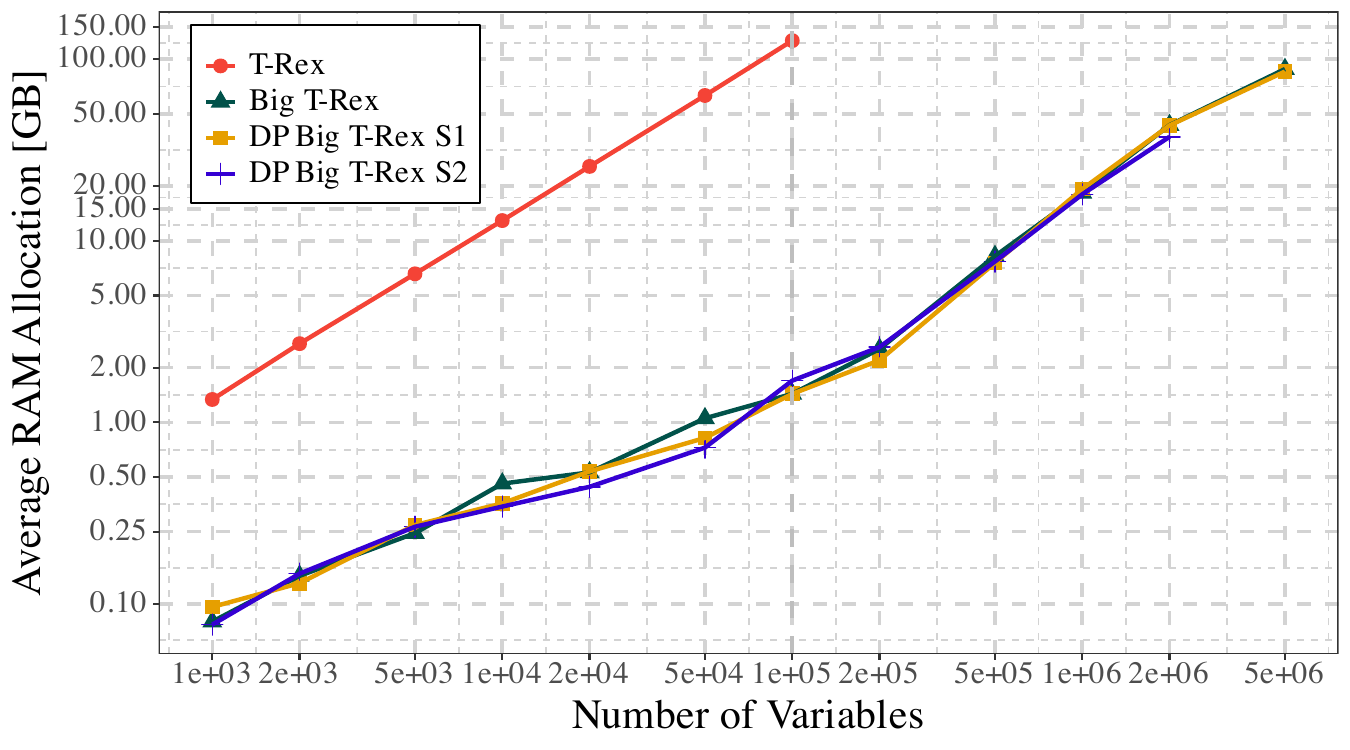}
    \caption{Average cumulative RAM allocations in GB for one \emph{T-Rex} run. 
             Note that both axes are scaled logarithmically.
             }
    \label{fig:Memory_Demand_TRexes}
\end{figure}

\begin{figure}[htp]
    \centering
    \includegraphics[width=0.85\linewidth]{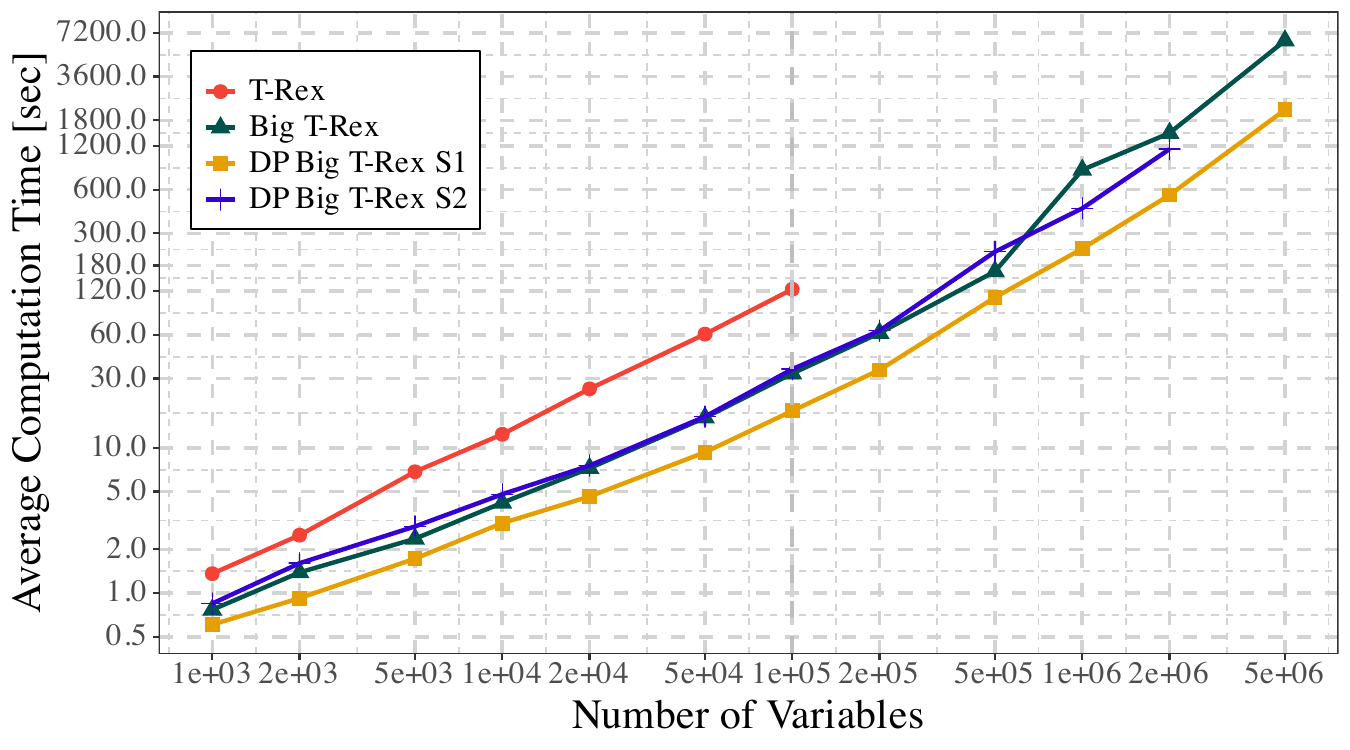}
    \caption{Average computation time for one \emph{T-Rex} run in seconds.
             Note that both axes are scaled logarithmically.
             }
    \label{fig:Time_Demand_TRexes}
\end{figure}

The second setup evaluates the performance in cumulative RAM allocations for one \emph{T-Rex} run and its computation time, for a varying set of predictor variables $p$.
We fixed $n = 300$ and $p \in \{1\mathrm{e}3, 2\mathrm{e}3, 5\mathrm{e}3, 1\mathrm{e}4, 2\mathrm{e}4, 5\mathrm{e}4, 1\mathrm{e}5, 2\mathrm{e}5, 5\mathrm{e}5, 1\mathrm{e}6, 2\mathrm{e}6, 5\mathrm{e}6\}$.
All implementations were evaluated and averaged based on $25$ Monte Carlo trials for all computationally feasible settings.
Additionally, $L = p$, $\mathrm{SNR} = 1$, and $\mathrm{tFDR} = 0.2$ were chosen.
Fig.~\ref{fig:Memory_Demand_TRexes} shows that \emph{Big T-Rex} and \emph{DP Big T-Rex} provide huge RAM savings.
Also, according to Fig.~\ref{fig:Time_Demand_TRexes}, \emph{DP Big T-Rex} $\mathrm{S}_{1}$ delivers the best computation time performance, due to permuted, and not generated, random dummies in $\widetilde{\boldsymbol{X}}_{(k)}$.

\vspace{-5 pt}

\section{Conclusions}
\label{sec:Conclusions}
This work introduced the \emph{Big T-Rex} selector, an implementation of the \emph{T-Rex} selector that operates on memory-mapped data, uses highly-optimized computations, and allows to reduce the set of stored dummy matrices to a single candidate matrix via permutation strategies.
Our numerical experiments demonstrated that the RAM allocations were reduced by a factor of up to 88, and the computation time was reduced by a factor of up to 6.
As a result of this work, researchers and practitioners without access to HPCs can now compute FDR-controlled high-dimensional variable selection tasks with 5,000,000 variables in 30 minutes on a laptop.

With its unmatched scalability, substantial run-time improvements, and proven FDR-control, the \emph{Big T-Rex} has the potential to enable advanced analysis of large-scale high-dimensional data and provide new insights in applications, such as in genomics and proteomics research.

\clearpage
\printbibliography
\end{document}